\documentstyle[12pt]{article}
\newlength{\mathspace}

\begin{document}
\setlength{\oddsidemargin}{0cm}
\setlength{\baselineskip}{7mm}
\setlength{\mathspace}{2.5mm}
\def\beq{\begin{equation}}
\def\enq{\end{equation}}
\def\integ{\oint {dz\over 2\pi i}}

\begin{titlepage}

    \begin{normalsize}
     \begin{flushright}
                 UT-Komaba/95-12 \\
                 December 1995 \\
     \end{flushright}
    \end{normalsize}
    \begin{LARGE}
       \vspace{1cm}
       \begin{center}
         {PHYSICAL SPECTRA IN STRING THEORIES\\
{\Large --- BRST Operators and Similarity Transformations ---
\footnote{Invited talk given at the ``International Symposium on BRS
Symmetry'' (Sep.~18--22, 1995, RIMS Kyoto).}}}
  \\
       \end{center}
    \end{LARGE}

  \vspace{5mm}

\begin{center}
           Mitsuhiro K{\sc ato}
           \footnote{E-mail address:
              kato@hep1.c.u-tokyo.ac.jp} \\
      \vspace{4mm}
        {\it Institute of Physics, University of Tokyo, Komaba} \\
        {\it Meguro-ku, Tokyo 153, Japan}\\
      \vspace{1cm}

    \begin{large} ABSTRACT \end{large}
        \par
\end{center}
\begin{quote}
 \begin{normalsize}
\ \ \ \
Several examples of similarity transformations
connecting two string theories with different backgrounds are reviewed.
We also discuss general structure behind the similarity transformations
from the point of view of the topological conformal algebra and of the
non-linear realization of gauge symmetry.

 \end{normalsize}
\end{quote}

\end{titlepage}
\vfil\eject

\setcounter{footnote}{0}
\section{Introduction}
In string theories, BRS symmetry plays crucial role especially in
defining physical spectrum\cite{F,KO,H,GSW}. Let us recall briefly
the basic ingredients for the BRS quantization of, say, bosonic
string.

Taking the conformal gauge, the BRS charge in left-moving
sector\footnote{We shall confine our argument in the left-moving
sector. The right-moving sector can be treated in exactly same way.}
is defined by
\beq
Q_B = \integ \left( c T + b c \partial c \right),
\enq
where $c(z)$ and $b(z)$ are the reparametrization ghost and
anti-ghost respectively, and $T(z)$ is the energy-momentum tensor for
the string coordinates (or the matter part, in the world-sheet
sense). $T(z)$ satisfies the OPE
\beq
T(y) T(z) \sim {c/2\over (y-z)^4} + {2\over (y-z)^2} T(z)
	+ {1\over y-z} \partial T(z),
\enq
where $c$ is the so-called central charge. The nilpotency of the BRS charge
$Q_B^2 = 0$ is guaranteed only when $c=26$\cite{KO}.

In use of the BRS charge physical states are defined by the
Kugo-Ojima\cite{KuOj} condition:
\beq
Q_B | phys \rangle = 0.
\enq
Solving this equation we obtain the physical spectrum of the theory
as BRS cohomology. For instance, the physical spectrum of the
critical bosonic string consists of the DDF\cite{DDF} states up to
BRS exact states\cite{KO,FO,Hen,FGZ}.

Thus $Q_B$ governs the physical spectrum of the string theory. If the
$Q_B$'s are related in some way between two apparently different
string theory, then the spectra of these theories should be also
related. Actually there exist such cases. In this talk I present some
examples in which the two different string theories are related by
the similarity transformations. Also I discuss about the meaning of
such kind of similarity transformations from the two points of view:
one is from topological algebra and the other is from non-linearly
realized gauge symmetry.

These observation will be important for the long-standing problem of
finding background independent formulation of string theory and
prospected universal theory of string. Even if we failed finding
them, it is still important for classifying the universality class of
string theories.

\section{2D black hole vs $c=1$ string}
Two-dimensional blackhole is constructed as a coset CFT of
$SL(2,R)/U(1)$\cite{W}. Here we skip how to define the theory and to derive
the metric, but just start from the current algebra. The $SL(2,R)$
current algebra is defined as
\begin{eqnarray}
J^0(y) J^0(z) &\sim& {-k/2\over (y-z)^2} ,\\
J^0(y) J^{\pm}(z) &\sim& {\pm\over y-z} J^{\pm}(z) ,\\
J^+(y) J^-(z) &\sim& {k\over (y-z)^2} - {2\over y-z} J^0(z) .
\end{eqnarray}
The energy-momentum tensor for $SL(2,R)$ part is given by the Sugawara
form:
\beq
T_{SL(2,R)} = {1\over k-2}\left[ {1\over 2}\left(J^+J^- +
J^-J^+\right) - J^0J^0 \right] ,
\enq
where the central charge for this $T_{SL(2,R)}$ is $c={3k\over k-2}$.
Gauging $U(1)$ is performed in standard way by first introducing the
gauge current $\tilde J(z)$ which satisfies
\beq
\tilde J(y)\tilde J(z) \sim {k/2\over (y-z)^2} ,
\enq
and anti-commuting ghosts $\xi(z)$ and $\eta(z)$. Then defining the
BRS charge for the diagonal part $J^0+\tilde J$
\beq
Q_{U(1)} = \integ  \eta(z) \left( J^0(z) + \tilde J(z) \right) ,
\enq
the coset $SL(2,R)/U(1)$ is given by the cohomology of $Q_{U(1)}$.

Now the total energy-momentum tensor for this system is given by
\beq
T_{\it tot} = T_{SL(2,R)} + {1\over k}\tilde J\tilde J - \xi\partial\eta ,
\enq
where the central charge is $c = {3k\over k-2} +1 -2$ which equals 26
if $k={9\over 4}$. According to the general procedure explained in
the introduction, BRS charge $Q_{\it diff}$ for the reparametrization
(or diffeomorphism) is constructed with this $T_{\it tot}$. Then the physical
spectrum is defined by the sum of both BRS charge of $U(1)$ and
diffeomorphism:
\begin{eqnarray}
& Q = Q_{\it diff} + Q_{U(1)} , \\
&Q | phys \rangle = 0 .
\end{eqnarray}

Let us consider the following free field representation of the
currents:
\begin{eqnarray}
J^0 &=& \sqrt{k\over 2} \partial u , \\
J^{\pm} &=& i\left(\sqrt{k'\over 2}\partial\phi\pm i\sqrt{k\over
2}\partial X\right)  e^{\pm i\sqrt{2\over k}(X+iu)} , \\
\tilde J &=& -i \sqrt{k\over 2} \partial v ,
\end{eqnarray}
where $k' = k-2$ and $\phi$, $X$, $u$ and $v$ are free boson fields.
With these fields $T_{\it tot}$ is rewritten as
\beq
T_{\it tot} = -{1\over
2}(\partial\phi)^2-{1\over\sqrt{2k'}}\partial^2\phi -{1\over
2}(\partial X)^2-{1\over 2}(\partial u)^2-{1\over 2}(\partial
v)^2-\xi\partial\eta ,
\enq
and BRS charge as
\beq
Q = \integ \left( cT_{\it tot}+bc\partial c \right) + \integ\sqrt{k\over
2}\eta(\partial u -i\partial v) .
\enq
Note that first three terms in the free field representation of
$T_{\it tot}$ are the same expression of the energy-momentum tensor
$T_{c=1}$ for flat $c=1$ string theory if we regard $X$ and $\phi$ as
matter and Liouville field respectively. In the BRS charge, however,
reparametrization ghost and $u$, $v$, $\xi$ and $\eta$ are not
decoupled, so we are not able to consider $c=1$ part and the rest
separately.

In Ref.\cite{IK1} we found the similarity transformation which brings
everything into decoupled form of $c=1$ and the rest. The
transformation is generated by the operator
\beq
R = \integ {1\over\sqrt{2k}} c \xi ( \partial u + i \partial v ) ,
\enq
with which the BRS charge is transformed as
\begin{eqnarray}
e^R Q e^{-R} &=& Q_{c=1} + Q_{U(1)}\\
&=& \integ \left( c T_{c=1} + bc\partial c \right) +
    \integ \sqrt{k\over 2} \eta (\partial u - i \partial v ) .
\end{eqnarray}
Here the BRS charge is decomposed into two totally decoupled parts:
one is for the $c=1$ string and the other is for topological sector
consists of $u$, $v$, $\xi$ and $\eta$. With this form BRS cohomology
becomes much simpler, i.e. just a direct product of independent
cohomologies of $Q_{c=1}$ and $Q_{U(1)}$. Moreover, the cohomology of
$Q_{U(1)}$ is trivial except for the zero mode of $\eta$; they are
topological.  Thus we have the total cohomology space
\beq
H^*_{SL(2,R)/U(1)} \simeq H^*_{c=1} \otimes H^*_{U(1)} = H^*_{c=1}
\oplus \eta_0 H^*_{c=1} .
\enq

The $c=1$ string lives essentially in the flat background while the
$SL(2,R)/U(1)$ string does in black hole, so this transformation
relates the two apparently different background.  Clarifying the
meaning of the transformation should give new light on the background
(in-\nolinebreak[4]) dependence of the string theory. We will try to
give some hints toward this in later sections.

\section{Polyakov's light-cone gauge}
Quite similar structure as we saw in the previous section exists also
in the relationship between conformal gauge and light-cone gauge {\it
a la\/} Polyakov\cite{P,KPZ} in the non-critical string theory. BRS
quantization in the light-cone gauge is first discussed in
Ref.\cite{I,K,Ho} and refined later\cite{MO} to accommodate the
so-called discrete states.

According to the Ref.\cite{I} we start from the level $k$ $SL(2,R)$
current algebra generated by $J^{\pm}$ and $J^0$. The Energy-momentum
tensor for the gravity sector is given by the Sugawara form with
improvement term:
\beq
T_{grav} = {1\over k-2} \left[{1\over 2}\left(J^+J^- + J^-J^+\right)
- J^0J^0 \right] + \partial J^0 .
\enq
We denote the energy-momentum tensor of the matter sector by $T_m$
with its central charge $c_m$. The total energy-momentum tensor $T =
T_m + T_{grav}$ and the current $J^+$ satisfy the following closed
algebra,
\begin{eqnarray}
T(y) J^+(z) &\sim& {1\over y-z}\partial J^+(z) , \\
J^+(y) J^+(z) &\sim& 0 , \\
T(y) T(z) &\sim& {c/2\over (y-z)^4} + {2\over (y-z)^2}T(z)
	{1\over y-z}\partial T(z) ,
\end{eqnarray}
where the central charge is given by $c = {3k\over k-2}+6k+c_m$.

Introducing ghost fields $b$, $c$ for the generator $T$ and $\xi$,
$\eta$ for $J^+$, BRS charge is defined by
\beq
Q_{l.c.} = \integ \left[c(T_m+T_{grav}+\partial\xi\eta)+bc\partial
c\right]
	+\integ\eta J^+ .
\enq
This time, we use Wakimoto's representation of $SL(2,R)$ current
\begin{eqnarray}
J^+ &=& \beta , \\
J^0 &=& \beta\gamma - \sqrt{k'\over 2}\partial\phi ,\\
J^- &=& \beta\gamma^2 - k\partial\gamma - 2\sqrt{k'\over
2}\partial\phi\gamma ,
\end{eqnarray}
where $\beta$ and $\gamma$ are commuting ghosts and $\phi$ is free
boson.
In use of these, $T_{grav}$ and $Q_{l.c.}$ are rewritten as
\begin{eqnarray}
T_{grav} &=& \partial\beta\gamma -{1\over 2}(\partial\phi)^2
	-{1\over 2}(\alpha_++{2\over\alpha_+})\partial^2\phi ,\\
Q_{l.c.} &=& \integ\left[c\left(T_m-{1\over 2}(\partial\phi)^2
	-{1\over 2}(\alpha_++{2\over\alpha_+})\partial^2\phi\right)
	+bc\partial c +c\partial\beta\gamma +c\partial\xi\eta\right]
	\nonumber\\
	& & +\integ\eta\beta ,
\end{eqnarray}
where $\alpha_+ = \sqrt{2\over k'}$.

Besides two terms $c\partial\beta\gamma +c\partial\xi\eta$,
$Q_{l.c.}$ is the sum of two independent parts, one of which is the
same expression of the BRS charge in the conformal gauge if we regard
$\phi$ as the Liouville field, and the other part is topological.
Again we have similarity transformation\cite{IK2} generated
by\footnote{This $R$ operator already appeared in \cite{I}, although
it was not recognized as the generator of the similarity
transformation.}
\beq
R = \integ\left(-\gamma c\partial\xi\right).
\enq
This eliminates just undesired terms and brings $Q_{l.c.}$ into the
sum of BRS charge of conformal gauge and that of topological model,
i.e. $e^R Q_{l.c.} e^{-R} = Q_{conf} + Q_{top}$.
This establishes the relationship of both gauges\footnote{See
\cite{Ta} for the treatment of the right-mover.}.

\section{Twisted $N=2$ SCA in the topological sector}
In the previous two examples, black hole and light-cone gauge, there
is a common feature, i.e. the existence of topological sector. We
have nice algebraic machinery in such topological systems: twisted
$N=2$ superconformal algebra
\begin{eqnarray}
G^{\pm}(y)G^{\pm}(z) &\sim& 0,\\
G^+(y)G^-(z) &\sim& {d/3\over (y-z)^3}+{1\over (y-z)^2}J(z)
	+{1\over y-z}T_{top}(z),\\
J(y)G^{\pm}(z) &\sim& {\pm 1\over y-z}G^{\pm}(z),\\
J(y)J(z) &\sim& {d/3\over (y-z)^2}.
\end{eqnarray}
Actually, these generators can be expressed by the commuting ghosts
$\beta$, $\gamma$ and the anti-commuting ghosts $\xi$, $\eta$ with a
parameter $\lambda$ as
\begin{eqnarray}
G^+ &=& \eta\beta ,\\
G^- &=& \lambda\partial\xi\gamma + (\lambda -1)\xi\partial\gamma ,\\
J   &=& \lambda\beta\gamma + (\lambda -1)\xi\eta ,\\
T_{top}&=& \lambda (\partial\beta\gamma + \partial\xi\eta )
	+ (\lambda -1)(\beta\partial\gamma + \xi\partial\eta).
\end{eqnarray}
where the central charge is given by $d = 3(1-2\lambda)$. The BRS
charge of this topological system is expressed in terms of twisted
$N=2$ generator as $Q_{top} = \integ G^+(z)$, so we can identify the
fields and the parameter $\lambda$ for the previous examples; for the
light-cone gauge case $\lambda = 1$ and fields are literal, while
black hole case $\lambda = 0$ and $\beta$, $\gamma$ are identified
with $\sqrt{k/2}\partial(u-iv)$, $1/\sqrt{2k}(u+iv)$ respectively.

In terms of this topological algebra the generator of similarity
transformation $R$ in both cases can be expressed in common way
simply as
\beq
R = \integ (-c(z)G^-(z)).\label{R}
\enq
Hence it turns out that the mechanism eliminating the coupling
terms comes from the topological algebraic origin.

Generally, if we have two sectors with energy-momentum tensor
$T_{c=26}$ and $T_{top}$ whose central charges are $c=26$ and $c=0$
respectively, and the $c=0$ part is governed by the twisted $N=2$ SCA
supplemented with the relation $G^-(y)T_{c=26}(z) \sim 0$, then BRS
charge for this string
\beq
Q=\integ\left[c(T_{c=26}+T_{top})+bc\partial c\right]+\integ G^+
\enq
is transformed into totally decoupled form
\beq
e^R Q e^{-R}=\integ\left[cT_{c=26}+bc\partial c\right]+\integ G^+
\enq
by the $R = \integ(-cG^-)$. We note here that the analogous structure
is known in the topological string case\cite{EKYY}.

\section{$G/H$ coset CFT coupled to 2D gravity}
The previous argument can be generalized to $G/H$ coset. Let us
denote $J^A(z)$ as a current of $G$ current algebra
\beq
J^A(y)J^B(z) \sim {k/2\over (y-z)^2}\delta^{AB} + {if^{ABC}_G\over
y-z}J^C(z),
\enq
with energy-momentum tensor
\beq
T_G(z) = {1\over k+h_G}J^A(z)J^A(z),
\enq
where $h_G$ is defined by the relation
$f_G^{ACD}f_G^{BCD}=h_G\delta^{AB}$.

Let $H^a(z)$ be a current associated with the subgroup $H$ of $G$,
for which energy-momentum tensor is denoted by $T_H(z)$. According to
the standard procedure, $G/H$ coset is constructed by gauging $H$
part. We introduce gauge current $\tilde H^a(z)$ which satisfies same
OPE as $H^a(z)$ but with
the level $\tilde k$ defined by the relation $k+\tilde k+2h_H=0$, and
a set of anti-commuting ghosts $\xi^a(z)$ and $\eta^a(z)$ with the
OPE $\xi^a(y)\eta^b(z) \sim {\delta^{ab}\over y-z}$. Then BRS charge
\beq
Q_H = \integ \left[ \eta^a(H^a+\tilde H^a)
	- {i\over 2}f_H^{abc}\xi^a\eta^b\eta^c\right]
\enq
defines the $G/H$ physical states. The total energy-momentum tensor
is a sum of each for $G$ current, $\tilde H$ current and ghosts:
$T_{total}=T_G+T_{\tilde H}-\xi^a\partial\eta^a$. This expression can
be rearranged into the sum of each for $G/H$ and $H/H$
\beq
T_{total}=T_{G/H}+T_{H/H} ,
\enq
where
\begin{eqnarray}
T_{G/H}&=&T_G-T_H ,\\
T_{H/H}&=&T_H+T_{\tilde H}-\xi^a\partial\eta^a .
\end{eqnarray}
The $H/H$ part is topological as $T_{H/H}$ is $Q_H$ exact\cite{KS}
\beq
T_{H/H}=\left\{ Q_H\, ,\, {1\over k+h_H}\xi^a(H^a-\tilde H^a)
\right\}.
\enq
As a string theory this $G/H$ matter couples to two-dimensional
gravity, so the total BRS charge is the sum of $Q_{\it diff}$ made of
$T_{total}$ and $Q_H$
\begin{eqnarray}
Q&=&Q_{\it diff}+Q_H\\
&=&\integ\left[cT_{total}+bc\partial c\right] + Q_H .
\end{eqnarray}

Again the similarity transformation can be defined as before.
That is to say the generator
\beq
R=\integ\,{-1\over k+h_H}c\xi^a(H^a-\tilde H^a)
\enq
transforms BRS charge as
\beq
e^RQe^{-R}=\integ\left[cT_{G/H}+bc\partial c\right] + Q_H ,
\enq
so that it is separated into $G/H$ string and $H/H$ topological parts.

At this point a natural question arises; are there twisted $N=2$ SCA
also for this system? The answer is no, instead we have topological
Kazama algebra\cite{Kaz}, which is realized as follows:
\begin{eqnarray}
G^+&=&\eta^a(H^a+\tilde H^a)-{i\over 2}f_H^{abc}\xi^a\eta^b\eta^c ,\\
G^-&=&{1\over k+h_H}\xi^a(H^a-\tilde H^a) ,\\
J&=&-\xi^a\eta^a ,\\
T_{top}&=&{1\over k+h_H}H^aH^a+{1\over\tilde k+h_H}\tilde H^a\tilde
H^a
	- \xi^a\partial\eta^a ,\\
F&=&{-1\over 2(k+h_H)^2}\left[h_H\xi^a\partial\xi^a
	+ if_H^{abc}\xi^a\xi^b(H^c+\tilde H^c)\right] ,\\
\Phi &=&{-1\over 6(k+h_H)^2}if_H^{abc}\xi^a\xi^b\xi^c .
\end{eqnarray}
This is essentially the same construction in Ref.\cite{IR}. Note that
if $H$ is abelian $F$ and $\Phi$ disappear so that the algebra
reduces to twisted N=2 SCA.

As in the twisted $N=2$ case, the similarity transformation generator
$R$ is
expressed as eq.(\ref{R}) in terms of $G^{-}(z)$ in the Kazama
algebra.
However, the OPE $G^{-}G^{-}$ does not vanish contrary to twisted
$N=2$ SCA, instead
\beq
G^{-}(y)G^{-}(z) \sim {-2\over y-z}F(z).
\enq
Nevertheless, the mechanism still works for the separation of
topological sector from the string theory. It is interesting to
clarify to what extent this structure can be generalized into more
general topological theory.\footnote{In course of the symposium,
it is pointed out by R.~Stora that our similarity transformation is
in parallel with the Kirkman map in the equivariant cohomology. I
thank R.~Stora for the comment.}

\section{$N=0$ string as $N=1$ string}
In this section, we describe another example which includes
supersymmetric generator. This is called Berkovits-Vafa\cite{BV}
superstring which is a special $N=1$ fermionic string equivalent to
$N=0$ string. In a sense, it gives an example of a certain vacuum of
$N=1$ string on which world sheet supersymmetry is broken to $N=0$.

We begin with an arbitrary $c=26$ energy-momentum tensor $T_m$ and
spin $(\frac{3}{2},-\frac{1}{2})$ fermionic ghosts $b_1$, $c_1$. They
form a $c=15$ $N=1$ SCA as follows:
\begin{eqnarray}
T_{N=1}&=&T_m-{3\over 2}b_1\partial c_1-{1\over 2}\partial b_1c_1
	+{1\over 2}\partial^2(c_1\partial c_1) ,\\
G_{N=1}&=&b_1+c_1(T_m+\partial c_1b_1)+{5\over 2}\partial^2c_1 .
\end{eqnarray}
They satisfy the OPE
\begin{eqnarray}
  T_{N=1}(y) T_{N=1}(z) & \sim & \frac{15/2}{(y-z)^4}
        + \frac{2}{(y-z)^2} T_{N=1}(z) + \frac{1}{y-z} \partial
T_{N=1}(z) ,\\
  T_{N=1}(y) G_{N=1}(z) & \sim & \frac{3/2}{(y-z)^2} G_{N=1}(z)
                        + \frac{1}{y-z} \partial G_{N=1}(z) ,\\
  G_{N=1}(y) G_{N=1}(z) & \sim & \frac{10}{(y-z)^3} +
       \frac{2}{y-z} T_{N=1}(z) .
\end{eqnarray}
BRS charge for this string is made up with these operators
\beq
Q_{N=1}=\integ\left(cT_{N=1}-{1\over 2}\gamma G_{N=1}+bc\partial c
	-{1\over 4}b\gamma^2+{1\over 2}\partial c\beta\gamma
	-c\beta\partial\gamma\right) \label{QN1} ,
\enq
where anti-commuting ghosts $b$, $c$ are associated with the generator
$T_{N=1}$ and commuting ghosts $\beta$, $\gamma$ with $G_{N=1}$.

It seems to be complicated to prove with the expression (\ref{QN1}) that
this system is actually equivalent to the $N=0$ string. The following
similarity transformation, however, makes things astonishingly
simple\cite{IK3}.
That is to say, with
\beq
R=\integ c_1\left({1\over 2}\gamma b-3\partial c\beta-2c\partial\beta
  -{1\over 2}\partial c_1cb+{1\over 4}\beta\gamma\partial c_1\right) ,
\label{RR}
\enq
BRS charge $Q_{N=1}$ is transformed into much simpler form: just a sum of
 each for $N=0$ string and topological system
\beq
e^R Q_{N=1} e^{-R} = Q_{N=0} + Q_{top} ,
\enq
where
\begin{eqnarray}
Q_{N=0}&=&\integ (cT_m+bc\partial c) ,\\
Q_{top}&=&\integ \left(-{1\over 2}b_1\gamma\right) .
\end{eqnarray}
Moreover, the cohomology of $Q_{top}$ is trivial, i.e. only a vacuum.
Thus the cohomology of $Q_{N=1}$ is isomorphic to that of $Q_{N=0}$.

In this case, the identification of twisted $N=2$ generator in the
operator $R$ is not clear at first sight. The similarity
transformation, however, can be decomposed into two steps one of which
is actually expressed by the twisted $N=2$ generator. Namely, as the first
step
\beq
R_{1}=\integ\left({1\over 2}c_{1}b\gamma
    -{1\over 2}c_{1}\partial c_{1}\beta\gamma\right)
\enq
transforms BRS charge into simpler form which resembles former
examples
\beq
e^{R_{1}}Q_{N=1}e^{-R_{1}}
=\integ\left[c(T_{m}+T_{b_{1}c_{1}\beta\gamma}+bc\partial c\right]
+\integ\left(-{1\over 2}b_{1}\gamma\right) .
\enq
Then we can identify twisted $N=2$ generators as $G^{+}=-{1\over
2}b_{1}\gamma$ and $G^{-}=c_{1}\partial\beta +3\partial c_{1}\beta$.
In terms of this, the second step is now familiar form
\beq
R_{2}=\integ (-cG^{-}) ,
\enq
\beq
e^{R_{2}}\left(e^{R_{1}}Q_{N=1}e^{-R_{1}}\right)e^{-R_{2}}
    =Q_{N=0}+Q_{top} .
\enq
Thus in this way we can see again the role of the twisted $N=2$
algebra in the similarity transformation.

\section{Non-linear realization}
So far we have looked at the similarity transformation from the
topological algebra point of view. There is another standpoint from
which we can reinterpret the transformation, i.e. the non-linear
realization of gauge symmetry.

Let us consider a finite dimensional Lie group $G$ for the illustration
of the idea\cite{Mc2}.  Let $T_a$ ($a=1,\dots ,\dim(G)$) be a generator
of $G$.  Also we denote a generator of subgroup $H$ by $T_i$
($i=1,\dots ,\dim(H)$) and that of $G/H$ coset by $X_{\alpha}$
($\alpha=1,\dots ,\dim(G)-\dim(H)$).  Here we assume that $G/H$ is
symmetric
\begin{eqnarray}
\left[\, T_i\, ,\, T_j\, \right] &=& f_{ij}{}^k T_k ,\\
\left[\, T_i\, , X_{\alpha} \right] &=& f_{i\alpha}{}^{\beta} X_{\beta} ,\\
\left[ X_{\alpha} , X_{\beta} \right] &=& f_{\alpha\beta}{}^k T_k .
\end{eqnarray}

$G$-algebra valued one-form
\beq
g^{-1}dg = \omega^a T_a \label{one-form}
\enq
 satisfies Maurer-Cartan equation
\beq
d\omega^a + {1\over 2}f_{bc}{}^a\omega^b\wedge\omega^c = 0 .
\enq
This can be expressed by another parametrization, e.g.  right coset
parametrization
\beq
g = e^{y^iT_i}e^{\xi^{\alpha}X_{\alpha}} .
\enq
Denoting $\phi^i$ as a one-form on $H$-orbit defined by
\beq
e^{-y^iT_i}de^{y^iT_i} = \phi^i(y)T_i \, ,
\enq
one-form (\ref{one-form}) is rewritten as
\begin{eqnarray}
g^{-1}dg &=&
e^{-\xi^{\alpha}X_{\alpha}}\phi^{i}T_{i}e^{\xi^{\alpha}X_{\alpha}}
      + e^{-\xi^{\alpha}X_{\alpha}}de^{\xi^{\alpha}X_{\alpha}} \\
&=& \phi^{i}(T_{i}-\xi^{\beta}f_{\beta i}{}^{\alpha}X_{\alpha}+\cdots)
      +d\xi^{\alpha}(X_{\alpha}-{1\over
      2}\xi^{\beta}f_{\beta\alpha}{}^{i}T_{i}+\cdots) .
\end{eqnarray}
Comparing this expression and (\ref{one-form}), we obtain the
transformation matrix $U(\xi)$ for basis change
\beq
(\omega^{i}\,\omega^{\alpha}) = (\phi^{j}\, d\xi^{\beta})\,
U^{-1}(\xi) \label{U-1} .
\enq

One the other hand, vector field $Y_{a}$ on $G$ satisfies
\beq
[Y_{a} , Y_{b}] = f_{ab}{}^{c}Y_{c} ,\qquad
\omega^{a}(Y_{b}) = \delta^{a}{}_{b} .
\enq
Corresponding to the right-coset parametrization, we also have vector
field $\eta_{i}$ on $H$-orbit
\beq
[\eta_{i} , \eta_{j}] = f_{ij}{}^{k}\eta_{k} ,\qquad
\phi^{i}(\eta_{j}) = \delta^{i}{}_{j} .
\enq
Then the basis change for the vector field is obtained by
\beq
\left(\begin{array}{c}{Y_{i}}\\{Y_{\alpha}}\end{array}\right) = U(\xi)
\left(\begin{array}{c}{\eta_{j}}\\{\partial\over\partial\xi^{\beta}}
\end{array}\right) .
\label{U}
\enq
This relation was used to reproduce the non-linearly realized
super-current $G_{N=1}$ in the previous section\cite{Ku,Mc}, where
anti-commuting field $b_{1}$ is nothing but the Nambu-Goldstone
fermion associated to the broken generator $G_{N=1}$.

BRS charge $Q_{G}$ is a corresponding object with the exterior
derivative on $G$
\beq d = \omega^{a}Y_{a} ,\enq
and defined by
\beq
Q_{G} = c^{a}\left(Y_{a}+{1\over 2}T_{a}^{\it gh}\right) ,
\enq
where $T_{a}^{\it gh} = - f_{ab}{}^{c}c^{b}b_{c}$, and the ghost
variables satisfy $\{ b_{a},c^{b}\} = \delta_{a}{}^{b}$.
For the right-coset basis $(\phi^{i}\, d\xi^{\alpha})$ the exterior
derivative is expressed as
\beq
d = \phi^{i}\eta_{i} +
d\xi^{\alpha}{\partial\over\partial\xi^{\alpha}} ,
\enq
hence the corresponding BRS charge
\beq
\tilde Q_{G} = \tilde c^{i}\left(\eta_{i}-{1\over
2}f_{ij}{}^{k}\tilde c^{j}\tilde b_{k}\right) + \tilde
c^{\alpha}{\partial\over\partial\xi^{\alpha}} ,
\enq
where the $\tilde b_{a}$ and $\tilde c^{a}$ are another set of ghosts
satisfies $\{ \tilde b_{a},\tilde c^{b}\} = \delta_{a}{}^{b}$.

$\tilde Q_{G}$ should be obtained by the basis change (\ref{U-1}) and
(\ref{U}) from $Q_{G}$. The basis change of ghost is thereby induced
\beq
c^{a}\, \rightarrow\, \tilde c^{a} = c^{b}(U^{-1})_{b}{}^{a} .
\enq
This can be achieved by the similarity transformation, namely
\beq
e^{R}c^{a}e^{-R} = c^{b}(U^{-1})_{b}{}^{a} ,
\enq
where the operator $R$ is defined by
\beq
R = c^{a}K_{a}{}^{b}b_{b} ,
\enq
such that $(e^{K})_{b}{}^{a} = (U^{-1})_{b}{}^{a}$.

These arguments can be generalized to the infinite dimensional case and
also the case which includes fermionic generators. Actually the
similarity transformation (\ref{RR}) in the previous section was able to be
reproduced in this way\cite{Mc2}.

Moreover, the transformations in the blackhole and light-cone gauge
examples can also be reproduced in this context.
For the 2D blackhole case, the starting algebra is a
closed algebra generated by the currents $T_{\it tot}$ and $J_{U(1)} =
J^{0} + \tilde J$. The $J_{U(1)}$ is non-linearly realized as
$\sqrt{k\over 2}\partial(u-iv)$ and so $u-iv$ is Nambu-Goldstone (N-G) boson.

Similarly, in the light-cone gauge case, the algebra is generated by
$T$ and $J^{+}$. And $J^{+} = \beta$ is broken generator, so that
the $\beta$ is N-G boson.

\section{Summary}
We have analyzed the various cases of the similarity transformations
in terms of the topological algebra and of the non-linear realization
of gauge symmetry.
We have shown the universal structure behind the similarity
transformation which may play an important role to understand the
background (in-)dependence of the string theory.

It is, in a sense, natural to be able to understand the same phenomena
from two different approaches, topological algebra and non-linear
realization; they both describe the decoupling of gauge degrees of freedom.

Recently, the notion of duality is being drawn much attention in
order to understand non-perturbative aspects of the string theory.
In this regards, the discussion extended here may shed another
light on the problems. For example, the T-duality can be understood
by the gauge symmetry\cite{Ev} which is always broken unless it is
on the self-dual point. Then our argument in the previous section
can be cast into the game. This will be reported elsewhere.

\section*{Acknowledgements}
The author would like to express sincere thanks to Hiroshi Ishikawa
for the collaboration in both published and unpublished works on which
the present talk is based.

\end{document}